\documentclass[%
 reprint,
 amsmath,amssymb,mathtools,
 aps,
pra,
]{revtex4-2}

\usepackage{graphicx}
\usepackage{subcaption}
\usepackage{dcolumn}
\usepackage{bm}
\usepackage{lipsum}
\usepackage{multirow}
\usepackage{xcolor}

\begin{document}

\preprint{APS/123-QED}

\title{Cold atom-ion systems in radiofrequency multipole traps: event-drive molecular dynamics and stochastic simulations }

\author{Mateo Londoño}
\author{Javier Madroñero}
\affiliation{%
Centre for Bioinformatics and Photonics (CIBioFi), Universidad del Valle, Edificio E20 No.~1069, 760032 Cali, Colombia}%
\author{Jesús Pérez-Ríos}%
\affiliation{Department of Physics and Astronomy, Stony Brook University, Stony Brook, NY 11794, USA 
}
\affiliation{Institute for Advanced Computational Science, Stony Brook University, Stony Brook, New York 11794, USA}

\date{\today}

\begin{abstract}
We have studied the general aspects of the dynamics of an ion trapped in an ideal multipolar radiofrequency trap while interacting with a dense cold atomic gas. In particular, we have explored the dynamical stability, the energy relaxation and the characteristic harmonic motion exhibited by a trapped Yb$^{+}$ ion in different multipolar potentials and immersed in various cold atomic samples (Li, Na, Rb, Yb). For this purpose, we used two different molecular dynamics simulations; one based on a time-event drive algorithm and the other based on the stochastic Langevin equation. Relevant values for experimental realizations, such as the associated ion's lifetimes and observable distributions, are presented along with some analytical expressions which relate the ion's dynamical properties with the trap parameters. 

\end{abstract}

--\maketitle


\section{\label{sec:introcduction} Introduction}
The realization of cold hybrid atom-ion systems has revolutionized the field of atomic, molecular, and optical physics, leading to a new venue for studying impurity physics \cite{SCHURER2017, MIDYA2016, ARDILA2021,Esben2021}, atom-ion collisions \cite{COTE20166,TOMZA2019,perezriosbook,perezrios_rydberg,JESUS2022} and quantum information sciences~\cite{MONROE2013, BRUZEWICZ2019}. However, most ion-atom systems require a time-dependent trap to hold the ion. At the same time, the ion is brought in contact with an atomic gas. As the atom approaches the ion, it is pulled away from the center of the trap, leading to the well-known micromotion heating \cite{CETINA2012, PINKAS2020, LONDONO2022}. This effect represents a problem for most applications in the cold and ultracold regime. For example, in quantum information sciences, micromotion heating can reduce the efficiency of sympathetic ion cooling, enhancing atom losses due to laser heating caused by successive gates \cite{HITE2017}. Similarly, in cold chemistry, the time-dependent trap induces long-lived ion-atom complexes that could potentially affect the stability of the ion~\cite{JESUS2023}.



One solution to curb micromotion heating is implementing radiofrequency higher-order multipole traps for ion confinement \cite{TOMZA2019,BASTIAN2016}. These traps create an almost boxlike trapping potential with a sizable flat potential or field-free region in the center, reducing the heating effects~\cite{Wester2009}. However, despite the advantage of lowering micromotion effects, the multipole trap has some weaknesses. For example, the trapping properties depend on the ion's average distance to the trap's center, leading to a stochastic stability parameter. Additionally, no analytical solution can be found to the equations of motion, and the numerical study of the collisional dynamics is very cumbersome~\cite{BASTIAN2016,NIRANJAN2021}. Generally, the trapping stability is characterized by a molecular dynamics approach, leading to a partial understanding of thermalized ions' energy distribution and position as a function of the trap's nature~\cite{BASTIAN2016,NIRANJAN2021}. On the other hand, a stochastic approach based on the Langevin equation has recently been developed for ions in a quadrupole trap~\cite{BLATT1986, LONDONO2022}. The time-continuous nature of the stochastic approach allows for describing the relaxation process of the ion or spectral composition of the motion and the time-dependence of the resulting ion's distributions,  not considered in molecular dynamics simulation~\cite{BASTIAN2016,NIRANJAN2021}. In addition, the stochastic formulation typically results in shorter simulation times than elaborate molecular dynamics simulations. For example, as the trap order increases, the free field region of potential reduces micromotion heating, and the stochastic Langevin simulation approach will become highly efficient for the ion dynamics in a buffer gas for any mass ratio.


This work theoretically explores the dynamics of a single ion in a multipole trap in contact with an atomic gas. To this end, we use two different simulation methods. One is based on a novel event-driven molecular dynamics simulation method via sampling collisional times from realistic atom-ion collisions. The other relies on the Markovian Langevin equation. The first is highly efficient for studying the stability of atomic mixtures of interest at different temperatures and initial conditions. In contrast, the second, cheaper computationally speaking, gives a precise understanding of the thermalization process. As a result, we can describe the ion stability in the discrete and continuous time domain, leading to new insights into the thermalization of the ion. The paper is divided as follows. Sections \ref{sec:stability} and \ref{sec:energy_dynamics} focus on the dynamics stabilities and energy distributions of the trapped ions using the event-drive molecular dynamics. Section \ref{sec:langevin_eq} is devoted to the ion's dynamics using the  Langevin equation, which is valid for multipolar traps where thermal behavior dominates the final distributions or, equivalently, where micromotion heating effects are reduced. Finally, Section \ref{sec:conclusions} summarizes the main results of the work and present some perspectives.

\section{Trap depth and dynamic stability for a single ion}
\label{sec:stability}

The dynamics of an ion in a radiofrequency (RF) multipole trap can be described using the adiabatic approximation (see appendix \ref{sec:appendixA}). Within this approximation, the slow secular motion is decoupled from the fast micromotion, and the stability of the ion is described with a single dynamic stability parameter \footnote{Note that there is an additional $n$ in the definition compared to the parameter $\eta$ in previous works~\cite{BASTIAN2016, DIETER1992}, this is a consequence of our definition of the parameter $q$ as an independent value of $n$.}:

\begin{equation}
\label{eq:dyn_est}
    \eta  = qn(n-1)\bigg( \frac{r}{r_{0}} \bigg)^{n-2} ,
\end{equation}
where $n$ is the trap order, $r_{0}$ is the radius of the trap and $q = \frac{2eU_{\rm{AC}}}{m_{\rm{ion}}r_{0}^{2}\Omega_{\rm{RF}}^{2}}$. In this equation, $U_{\rm{AC}}$ stands for the voltage on the electrodes, $e$ is the atomic ion charge, $m_{\rm{ion}}$ represents the ion's mass and $\Omega_{\rm{RF}}$ is the trap frequency. Therefore, the dynamic stability for a single ion in a trap with n$>$2 depends on the distance of the ion to the center of the trap, becoming a stochastic variable due to collisions with the gas. As a result, for a given maximum value of the dynamic stability parameter, $\eta_{\text{max}}$, there is always a distance of the ion at which the dynamics become unstable, given by

\begin{equation}
\label{eq:R_cri}
    r_{\rm{cri}} = r_{0}\Bigg( \frac{\eta_{\rm{max}}}{qn(n-1)} \Bigg)^{1/(n-2)},
\end{equation}
where generally $\eta_{\rm{max}}=0.33$~\cite{DIETER1992}.

\begin{figure}[h]
\centering
\includegraphics[width=\columnwidth]{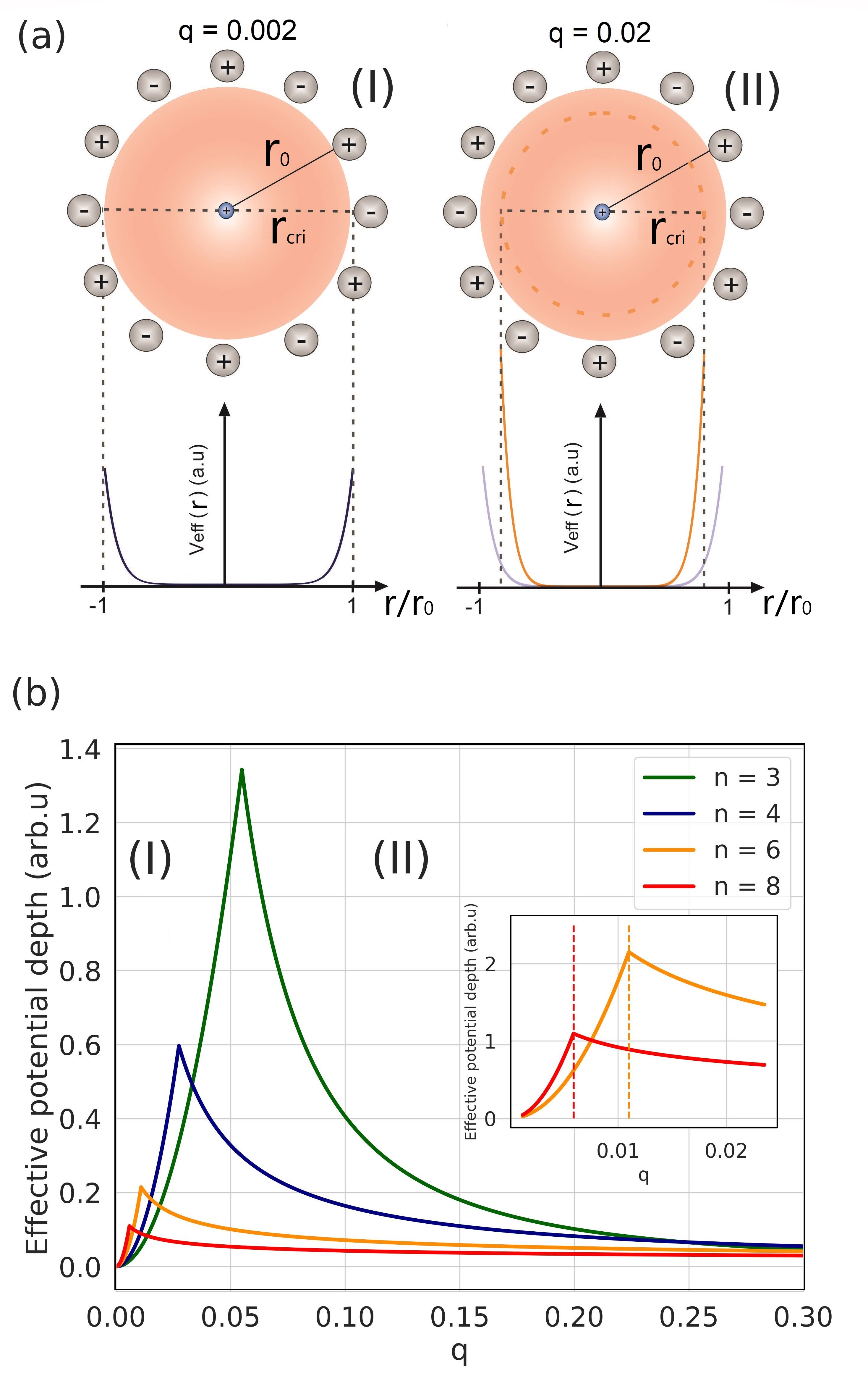}
\caption{The dependence of the effective trapping potential with respect to the trap parameters is presented. The panel (a) shows it schematically for a dodecapolar trap. The red cloud represents the homogeneous atomic cloud. For different values of the $q$-parameter the critical radius r$_{\rm{cri}}$ can be larger (region \textbf{(I)}) or shorter (region \textbf{(II)}) than the trap radius. Panel (b) shows the variation of the effective potential depth as a function of the $q$-parameter for a Yb$^{+}$ in different multipolar traps with a fixed RF-frequency of $\Omega_{\rm{RF}} = 2\pi$ MHz. Peaks of the effective depth define the transition between the regime where $r_{\rm{cri}} < r_{0}$ (\textbf{(I)}) and (\textbf{(I)}), and allow us to define a region of $q$ where ion dynamics is more confined and stable}
\label{fig:eff_depth_vs_q}
\end{figure} 

Next, following Refs.~\cite{MIKOSCH2007, MIKOSCH2008}, we introduce the trapping volume as the region where no effective energy transfer occurs between the ion and the field, also known as the field-free region. This region is bounded by $r_{\rm{cri}}$ when $r_{\rm{cri}} < r_{0}$, or by $r_{0}$ on the contrary, as illustrated in panel (a) Fig.\ref{fig:eff_depth_vs_q}. Hence, as the figure suggests, the effective trapping depth is a function of the RF frequency and the $q$-parameter, given by
\begin{equation}
\label{eq3}
V_{\rm{depth}}(\Omega;q) = m_{\rm{ion}} \frac{(qn\Omega)^{2}}{16}\frac{r_{\rm{tr}}^{2n-2}}{r_{0}^{2n-4}},
\end{equation}
where $r_{\rm{tr}} = \text{min} (r_{\rm{cri}}, r_{0})$. 

Due to the lack of stability diagrams in multipole traps, it is preferable to take the effective depth of the trapping potential as the reference to define stable trap parameters. Panel (b) of Fig. \ref{fig:eff_depth_vs_q} displays the effective depth of the trapping potential as a function of $q$ for $\Omega_{\rm{RF}} = 2\pi$ MHz, showing a maximum depth at r$_{\rm{cri}}$ = r$_{0}$, as expected in virtue of Eq.~(\ref{eq3}). Similarly, we notice that low-order traps show a larger trap depth than higher-order traps, indicative of the stronger stability of low-order traps versus high-order ones. As a result, for a given $\Omega_{\rm{RF}}$, we chose the initial dynamical stability $\eta$ based on the $q$-value where the maximum trap depth is observed. However, this choice will not guarantee the stability of the ion due to the inherent position-dependent stability. Hence, an approach based on the dynamics of the ion is required. 

\section{Event-driven molecular dynamics simulations}

Event-driven molecular dynamics simulations generalize the simulation approach of Zipkes \textit{et al.}~\cite{Zipkes2010} to the case of multipole traps. In our approach, we assume instantaneous hard-sphere collisions with an energy-independent scattering rate proper of the Langevin model for ion-atom collisions. However, because of the lack of analytical expressions for the ion's motion in a multipolar potential, we propagate the equation of motion from one collision event to the next one. Additionally, we consider a homogeneous atomic density. The event-drive algorithm consists of the following steps:

\begin{itemize}

\item \textit{Initialization:} We initialize the trap parameters ($\Omega_{\rm{RF}}$, $q$, $r_0$), including $n$, the temperature and density of the atomic cloud (T, $\rho$), and the ion initial conditions. Typically, we place the ion at a distance of 0.01~r$_{0}$ at rest, \textbf{v}$_{\text{ion},0}=0$.


\item \textit{Event time: } Once the initial conditions for the experiment are set up, we compute the event-time (t$_{c}$), associated with an atom-ion collision. This time is sampled from a Poisson distribution with mean value  $\tau = 1/\Gamma_{\rm{Lang.}}$, where the Langevin scattering rate $\Gamma_{\rm{Lang.}}$ depends on the atom-ion long range coefficient $C_{\rm{4}}$ and the atom-ion reduced mass $\mu$ as (in atomic units)
\begin{equation}
    \Gamma_{\rm{Lang.}} = 2\pi\sqrt{\frac{C_{\rm{4}}}{\mu}}.
\end{equation}

\item \textit{Preparing the collision:} With the event-time computed, we propagate the ion in the trap from the initial condition to the collision time. The propagation is carried out using a fourth order Runge-Kutta method to integrate the ion's equation of motion

\begin{equation}
\label{eq:ion_EOM}
    m_{\rm{ion}}\frac{d^{2}\textbf{r}(t)}{dt^{2}} = \textbf{F}(\textbf{r}) = \frac{eU_{\rm{AC}}n}{r_{0}^{n}}\cos(\Omega_{\rm{RF}}t)r^{n-1}\textbf{e}_{\rm{RF}},
\end{equation} 
where $\textbf{e}_{\rm{RF}}$ represents the unitary vector of the trapping force, which depends on the azimuthal angle $\phi$. At the same time, we pick up an atom from the ensemble. The velocity for the atom is sampled from a Maxwell-Boltzmann distribution associated to the temperature $T$ of the gas. If the final position of the ion's satisfy r$_{\rm{ion}} <$ r$_{\rm{0}}$ the collision is produced, if not the propagation will broke up and the ion is lost. 

\item \textit{Hard-sphere collision:} If the collision takes place, the ion's velocity changes following a hard-sphere collision with the atom, going from the initial value \textbf{v}$_{\text{ion},0}$ to 
\begin{equation*}
    \textbf{v}_{\rm{ion},f} = (1-\beta)\textbf{v}_{\rm{ion},0} + \beta\mathcal{R}(\theta, \phi)\textbf{v}_{\rm{ion},0},
\end{equation*}
with $\beta = \frac{1}{1+\zeta}$, $\zeta = m_{\rm{atom}}/m_{\rm{ion}}$, and $\mathcal{R}(\theta, \phi)$ represents the rotation matrix depending on the collision angles $\theta$ and $\phi$, which are sampled homogeneously from [0,2$\pi$], following Langevin cross section. The position of the ion remains unchanged.

\item \textit{Saving the observables:} For the energy of the ion we compute the secular velocity given by 
\begin{equation*}
    \textbf{v}_{\rm{sec},f} = (1-\beta)\textbf{v}_{\rm{sec},0} + \beta\mathcal{R}\textbf{v}_{\rm{sec},0} + \beta(\mathcal{R}-\textbf{1})\textbf{v}_{\rm{mm}}.
\end{equation*} 
The micromotion component $\textbf{v}_{mm}$ is computed using
\begin{equation}
\begin{split}
    \textbf{v}_{\text{mm}}
    &\approx -\frac{qn\Omega_{\rm{RF}}}{2}\sin(\Omega_{\rm{RF}} t)r^{n-1}_{\rm{sec}}\textbf{e}_{\rm{RF}},
\end{split}
\end{equation}
and using $r_{\rm{sec}} = r_{\rm{ion}}$ the ion's position at the end of the propagation. Then, the new velocity and position are settled as the initial conditions and we loop back to compute the event-time. 

\end{itemize}

The algorithm will finish when reaching the total number of collisions or when the ion gets lost from the trap. 

\section{Stability of a single ion in a multipole trap}

We explore the stability of a Yb$^{+}$ ion immersed in different cold atomic baths using the event-driven molecular dynamic simulation approach. In particular, we are interested in some atomic species previously studied in the context of cold collisions \cite{ZHANG2009, MCLAUGHIN2014, JOGER2017, PETROV2016}. In this study, stability is given as the survival probability of the ion one reaches thermalization with the bath. Specifically, for each atom-ion mixture, we simulate 1000 sympathetic cooling experiments. Once the ion thermalizes, we let the ion evolve for 500 extra collisions with the bath atoms. Once finished the simulations, stability is defined as the number of survival trapped ions (N$_{\rm{s}}$) over the number of experiments (N$_{\rm{Tot}}$). 

\begin{figure}[h]
\centering
\includegraphics[width=\columnwidth]{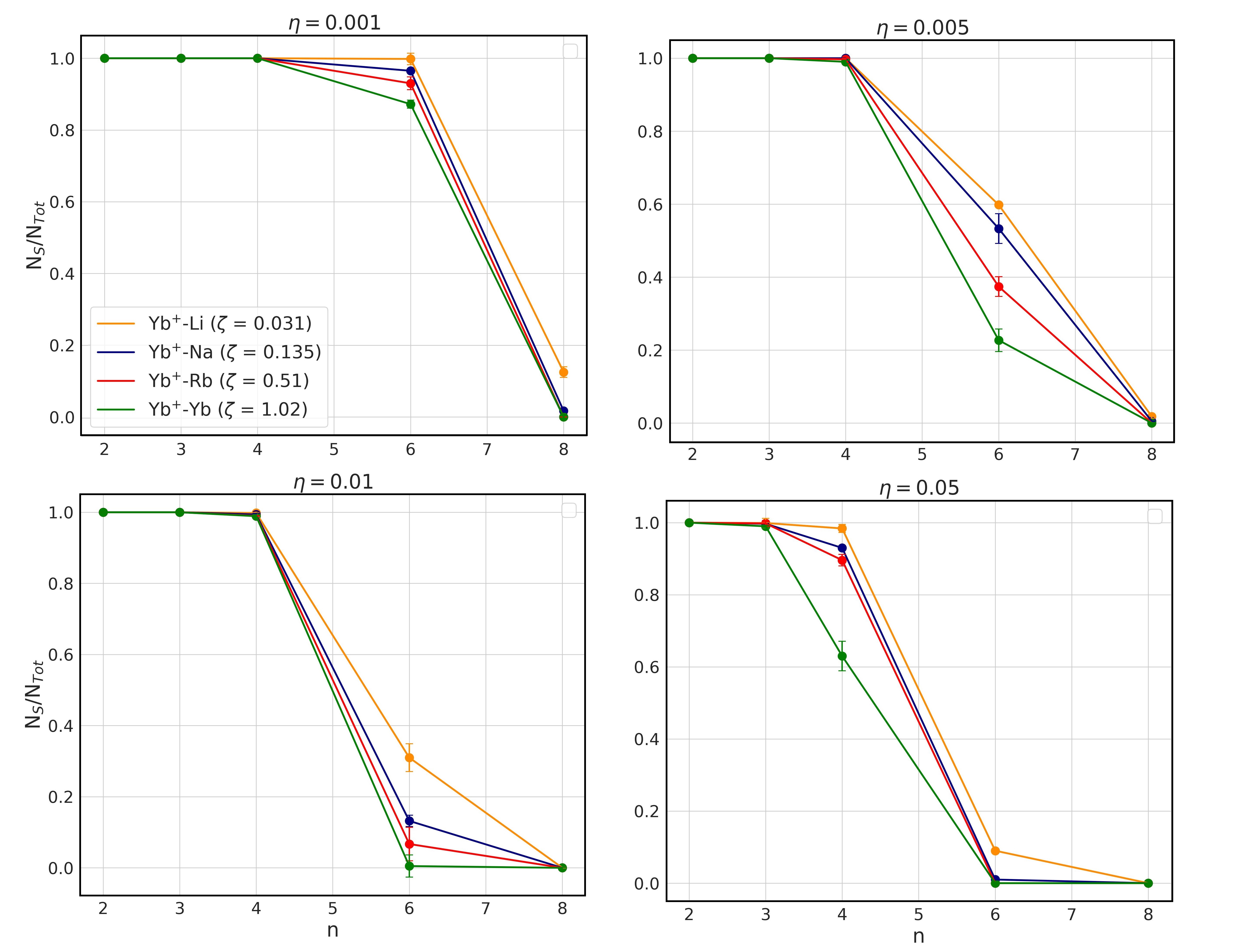}
\caption{Stability as a function of the trap order, for different atomic species and initial dynamical stabilities. As mentioned in the text we have defined stability as the survival probability of the ions at 500 collisions after thermalization. Because of the statistical nature of our simulation, we show the deviation as bar errors. All the simulations were performed at $T$=1$\times$10$^{-3}$K and $\rho$=1$\times$10$^{18}$m$^{-3}$ }
\label{fig:stabilities_and_trap_order}
\end{figure}

The results for our simulations are shown in Fig.~\ref{fig:stabilities_and_trap_order}, where the survival probability of the ion as a function of the trap order and initial $\eta$ value is displayed. All the simulations are carried out at $T$=1$\times$10$^{-3}$K and a sample density of 
$\rho$=1$\times$10$^{18}$m$^{-3}$. As the Figure shows, low-order traps represent high stability but are prone to present RF heating. Therefore, a dodecapolar trap is the best choice based on stability and RF heating. Furthermore, we have computed the average number of collisions before abandon the trap and its lifetime, as displayed in Table \ref{table:life_times_n6}. This Table presents the lifetimes and the mean number of collisions for some of the previously explored unstable configurations. There is a remarkable variance associated with the lifetimes, showing the high sensibility of the system to initial conditions as well as initial collisional events. The thermalization rate, reported for each mixture at different temperatures, is not affected by the trap properties and is reported for each mixture. In general, from the data, it can be seen that most of the systems reach thermalization before the ion leaves the trap.


\begin{figure}[]
\centering
\includegraphics[width=\columnwidth]{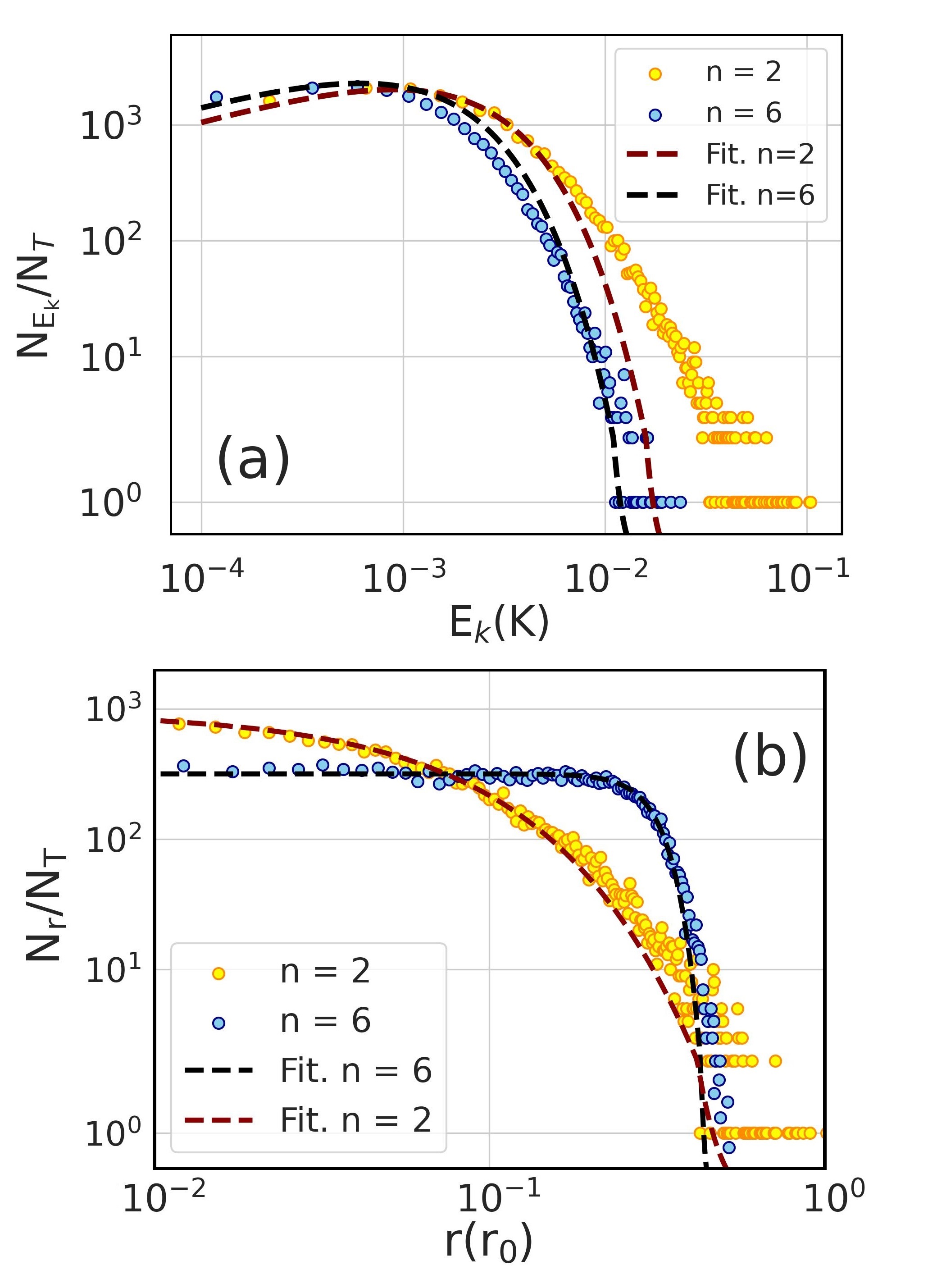}
\caption{Distributions for a Yb$^{+}$ in two different multipolar traps and in the presence of an Rb cold atomic gas. Panel (a) shows the kinetic energy distributions for the ion in a quadrupolar (n = 2) and dodecapolar (n = 6) trap, and the corresponding thermal fitting, the mean value of the energy depends on the trap order as explained in Sec. \ref{sec:energy_dynamics}. Panel (b) shows the position distributions for the same conditions. In both cases, it is observed how the high multipole trap leads to a thermal behavior for the same atom-ion mixture.}
\label{fig:kinetic_energy_poles}
\end{figure}

\begin{table*}[ht]
\centering
\begin{tabular}{|l|l|l|l|l|l|l|l|}
\hline
 &
\multicolumn{4}{|c|}{n=6} &
\multicolumn{2}{|c|}{n=8}&
No trap
\\
\cline{2-8}
Atom & \multicolumn{2}{c|}{$\eta = 0.005$ } & \multicolumn{2}{c|}{$\eta = 0.01$ } & \multicolumn{2}{c|}{$\eta = 0.005$ } &\\
\cline{2-7}
& $\,\,\,\,\,\, \tau$($\times 10^{-3}$s) & \,\, $N_{\rm{coll}} $ & $\,\,\,\,\,\,\,\, \tau$($\times 10^{-3}$s) & \,\,\,\,\, $ N_{\rm{coll}} $ \,\,\,\,\, & $\,\,\,\,\,\,\,\, \tau$($\times 10^{-4}$s) & \,\,\,\, $N_{\rm{coll}}$ 
& \,\,\,\, N$_{\rm{coll}}^{\rm{T}}$
\\ 
\hline \hline
Li  &  \,\,\, 3.5 $\pm$ 2.1 & \,\, 167 & \,\,\, 2.8 $\pm$ 1.8  & \,\,\,\,\,\, 136 &  \,\,\, 10.8 $\pm$ 8.1 & \,\,\,\, 57& \,\,\,\,27 \\  \cline{1-8}
Na & \,\,\, 6.8 $\pm$ 4.2 &\,\, 154 & \,\,\, 4.2 $\pm$ 3.0 & \,\, \,\,\,\,111 & \,\,\, 8.1 $\pm$ 5.4 &\,\,\,\, 22& \,\,\,\, 8 \\ \cline{1-8}
Rb& \,\,\, 6.1 $\pm$ 4.0 & \,\, 141 & \,\,\, 3.4 $\pm$ 2.1& \,\,\,\,\,\, 73  & \,\,\, 5.9 $\pm$ 4.2 & \,\,\,\, 14& \,\,   \\\cline{1-7}
Yb &\,\,\,  10.8 $\pm$ 8.2 & \,\, 124 & \,\,\, 5.1 $\pm$ 3.8 &\,\,\,\,\,\, 59 &\,\,\,  9.6 $\pm$ 7.5 & \,\,\,\, 11& \,\,\,\, $<$5 \\ \cline{2-8}
 \cline{1-4}
\end{tabular}
\caption{Mean lifetime and the number of collisions for the ion in a multipole trap with n=6 and n=8 at a temperature of $T$ = 1$\times$10$^{-3}$k and gas density $\rho$ = 1$\times$10$^{18}$m$^{-3}$. Two different initial stability parameters are displayed for the n=6 case, while only one is shown for n=8, given its low stability. The mean number of collisions for thermalization  (N$_{\rm{coll}}^{\rm{T}}$) is also reported}
\label{table:life_times_n6}
\end{table*}

Based on our study, ion losses usually take place after thermalization. Hence, the time-average probability of losses after thermalization should satisfy
\begin{equation}
  \bar{P}_{\rm{loss}} \propto  \bar{P}(r|r_{\rm{cri}})\tilde{P}(v|v_{\rm{cri}}),
\end{equation}
where  $\bar{P}(r|r_{\rm{cri}})$ is the time-average probability of finding the ion at the critical position given by Eq.~(\ref{eq:R_cri}) and  $\tilde{P}(v|v_{\rm{cri}})$ is the probability of having the ion with a velocity larger than a critical value $v_{cri}$. In general, the form of the distributions depends on the multipolar order, for low-order traps the micromotion heating leads to long-tail behavior which is notorious as the atom-to-ion mass ratio increases. Then, Tsallis-type distribution captures properly this behavior \cite{LONDONO2022, BASTIAN2016}. However, for trap-order n$>$4, which we are more interested here, the distributions tend to a thermal behavior as shown in Fig.~\ref{fig:kinetic_energy_poles}. Then, the average position distribution ($\bar{P}(r)$) satisfies
\begin{equation}
\label{eq:spatial_distribution}
  \bar{P}(r) \propto \exp (-V_{\rm{eff}}(r)/k_{\rm{B}}T),
  \end{equation}
going from a Gaussian-type distribution when $n$ = 2 to a box-like homogeneous distribution as the polar order tends to infinity. This box-like distribution tendency, displayed in panel (b) of Fig.~(\ref{fig:kinetic_energy_poles}), gives rise to a free-field region where the ion can move almost freely. As a result, in higher-order multipole traps, the ion is less localized, reaching more considerable distances from the trap's center, eventually leading to ion losses. 
However, the instability depends on the velocity of the ion at the critical distance too.

\begin{figure}[h]
\centering
\includegraphics[width=\columnwidth]{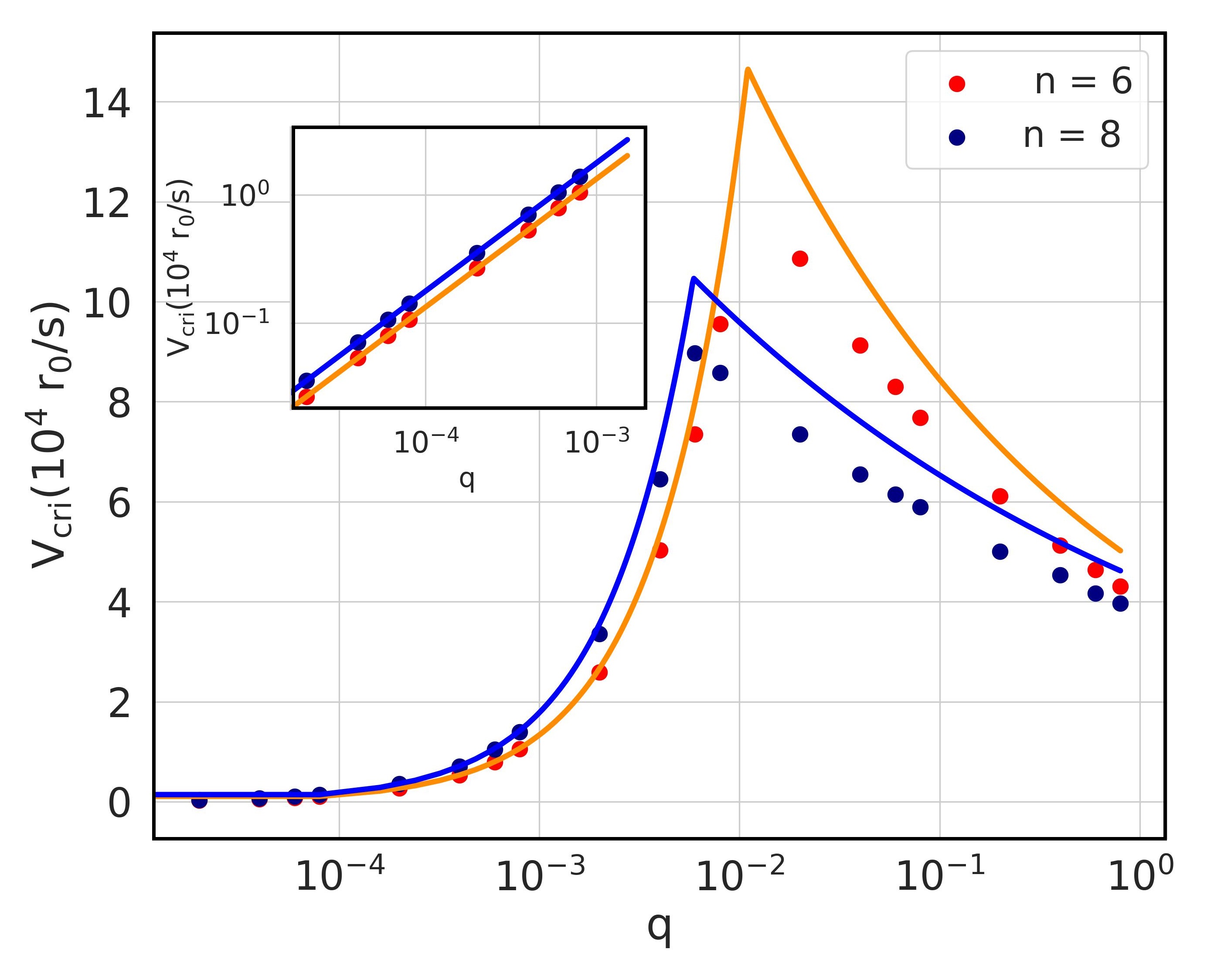}
\caption{Ion's critical velocity for two different trap orders. The dots correspond to numerical simulation and lines are the fitting by Eq.~(\ref{eq:v_cri}). An overestimation of the analytical model results from not considering micromotion and collision effects at the boundary of the field }
\label{fig:critical_velocity}
\end{figure}

$\bar{P}(v|v_{\rm{cri}})$ stands for the probability that the ion shows a larger velocity than a critical value $v_{\rm{cri}}$: the minimum value for having ion losses at the critical position and depends on the trap parameters. Assuming n$>$4, we can approximate the distribution to a thermal form as
\begin{equation}
\begin{split}
    \bar{P}(v|v_{\rm{cri}})  \propto \Theta(v-v_{\rm{cri}})v e^{\frac{-mv^{2}}{k_{\rm{B}}T}},
\end{split}    
\end{equation}
where $\Theta(x)$ is the Heaviside function of argument $x$, and $v_{\rm{cri}}$ depends on the trap properties. In general, $v_{\rm{cri}}$ is influenced by micromotion effects, as well as by possible collisions at the boundary where the field rises to its highest value. However, an estimate could be determined when the ion's kinetic energy at the boundary is equal to the effective potential depth,
\begin{equation}
    V_{\rm{depth}} (\Omega_{\rm{RF}}; q)  = m_{\rm{ion}}\frac{(qn\Omega_{\rm{RF}})^{2}}{16}\frac{r_{\rm{tr}}^{2n-2}}{r_{0}^{2n-4}} = \frac{1}{2} m_{\rm{ion}} v_{\rm{cri}}^{2}, 
\end{equation}
which leads to the following relations
\begin{equation}
     v_{\rm{cri}}(n;q) = \begin{cases} \frac{ n\Omega_{\rm{RF}}r_{0}}{\sqrt{8}}\,q, & \text{if } r_{0} \leq r_{\rm{cri}}, \\ \frac{ n\Omega_{\rm{RF}}r_{0}}{\sqrt{8}}\big( \frac{\eta_{max}}{n(n-1)}\big)^{\frac{2n-2}{2n-4}}\, q^{\frac{1}{2-n}}, & \text{if } r_{\rm{cri}} \leq r_{0}.\end{cases} .
     \label{eq:v_cri}
\end{equation}
Fig.~\ref{fig:critical_velocity} shows  $v_{\rm{cri}}$ for different multipole traps, comparing the full numerical approach based on event-driven molecular dynamics against our analytical results of Eq.~(\ref{eq:v_cri}). As a result, it is observed that for increasing values of $q$,  Eq.~(\ref{eq:v_cri}) overestimates the value of $v_{\rm{cri}}$ because it ignores micromotion and collision effects that could increase the secular velocity at the limit distance, producing losses for lower velocities than predicted. However, the model gives good agreement for low values of $q$ and an adequate qualitative description of the loss dynamics. Hence, as the temperature decreases, the stability will increase following the velocity distribution for a fixed trap configuration. Once the temperature is such that the critical velocity is not reached, the ion can only be lost through collision events at the boundary.

\section{Mean kinetic energy of an ion }
\label{sec:energy_dynamics}

In the adiabatic approximation, the virial theorem
\begin{equation*}
 \frac{2}{3}\langle E_{k}\rangle = (n-1)\langle V_{\rm{eff}} \rangle,
\end{equation*}
holds, resulting in the ratio $\langle V_{\rm{eff}} \rangle$/$\langle E_{k}\rangle$ = 2/3($n$-1), and the ion's mean kinetic energy as
\begin{equation}
\label{eq:expected_energy_no_mass}
    \langle E \rangle = \bigg[\frac{3}{2} + \frac{1}{n-1}\bigg]k_{\rm{B}}T
\end{equation}

However, it is possible to include mass effects assuming that micromotion degrees of freedom is assigned to the atom dynamics~\cite{BASTIAN2016}. We can then arrive at the expected value for the atomic energy
\begin{equation}
\label{eq:expected_energy_mass}
    \langle E_{a} \rangle = \frac{3}{2} k_{\rm{B}}T + \zeta V_{eff}(r),
\end{equation}
where $\zeta = m_{\rm{atom}}/m_{\rm{ion}}$ is the mass ratio. Combining Eqs.~(\ref{eq:expected_energy_no_mass}) and (\ref{eq:expected_energy_mass}) we propose that the mean kinetic energy of the ion can be described as
\begin{equation}
\label{eq:expected_energy_total}
    \langle E \rangle = \frac{3}{2}k_{\rm{B}}T + \frac{k_{\rm{B}}T}{n-1} + \frac{\alpha(n,\zeta)\zeta}{n-1} k_{\rm{B}}T,
\end{equation}
where $\alpha(n,\zeta)$ is a free parameter to determine, depending on the trap order and mass ratio.



\begin{figure}[]
\centering
\hspace{-0.1cm}
\includegraphics[scale=0.26
]{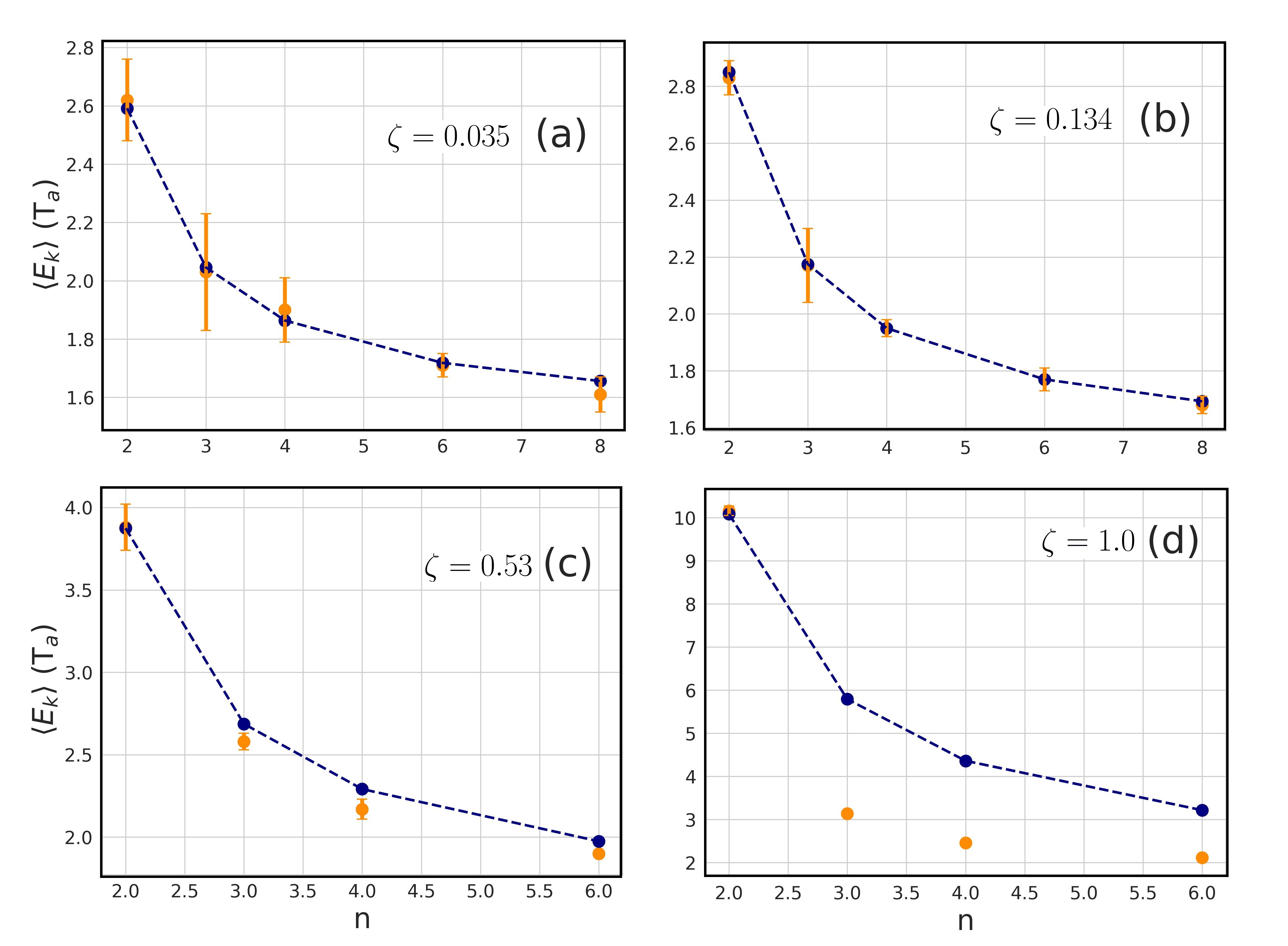}
\caption{Ion's energy for different trap order n. The fit curve is given by the expression \ref{eq:expected_energy_total} and is more accurate for low atom-to-ion mass ratios.  }
\label{fig:Energy_variation}
\end{figure}

Fig. \ref{fig:Energy_variation} displays the results for the mean energy of a $^{+}$Yb ion trap in a multipole trap in the presence of different atomic baths. For low mass ratio values, the required fitting parameter from Eq.~(\ref{eq:expected_energy_total}) is independent of the trap order. However, when the mass ratio approaches one, the trap order has a strong effect on the mean kinetic energy of the ion. In that case, Eq.~(\ref{eq:expected_energy_total}) is still applicable but with a fitting parameter for each trap order. 


From Eq.~(\ref{eq:expected_energy_total}), it is possible to identify the non-thermal component of the mean kinetic of the ion as 
\begin{equation}
    \Delta E = \frac{k_{\rm{B}}T}{n-1} + \frac{\alpha(n,\zeta)\zeta}{n-1} k_{\rm{B}}T.
\end{equation}
When this term is small compared to the thermal component of the ion kinetic energy, it is possible to use a temperature to describe the ion rather than its mean kinetic energy. Furthermore, note that Eq.~(\ref{eq:expected_energy_total}) allows expressing those contributions of the effective potential in Eq.~(\ref{eq:expected_energy_mass}) as thermal contributions, where the trap-parameters modifications are only incorporated in the fitting parameter $\alpha$, which can be easily computed.

\section{Langevin equation model}
\label{sec:langevin_eq}

The dynamics of a trapped ion in a neutral sea can be described by solving the Langevin stochastic equation of motion~\cite{LONDONO2022}. In this approach, all degrees of freedom of the bath are substituted by an effective stochastic force modeled by a Gaussian white noise $\mathbf{\zeta}$(t), whose components satisfy
\begin{equation}\label{WN}
\langle \zeta_{j}(t) \rangle = 0 \,\,\,\,\,\, \text{and} \,\,\,\,\,\, \langle \zeta_{j}(t)\zeta_{i}(s) \rangle = D\delta_{ij}\delta(t-s).
\end{equation}
Where $D$ is the diffusion coefficient related to the friction coefficient $\gamma$ by means of the fluctuation-dissipation theorem, $ \gamma = \frac{D}{2k_{B}T}$. Therefore, the diffusion and friction coefficients are related at a given temperature $T$. The friction coefficient, $\gamma$, encapsulates details of the atom-ion scattering through the thermally averaged diffusion cross-section, following the Chapman-Enskog approximation~\cite{LONDONO2022}.


Here, the stochastic equations of motion are formulated in Cartesian coordinates, using a multipolar expansion based on the analytical function $z = (x + \text{i} y)^{n}$ to obtain the RF field for a given trap order, $n$. The explicit derivation is shown in appendix \ref{sec:appendixB}. The equations of motion for each component of the ion's radial position, $r_{j}$, is given by 
\begin{equation}
\label{eq:langevin_eq}
\frac{d^{2}r_{j}}{dt^{2}} + \frac{\gamma}{m_{\rm{ion}}}\frac{dr_{j}}{dt}+ \frac{\Omega^{2}_{\rm{RF}}q_{j}}{2}\cos(\Omega_{\rm{RF}}t)\frac{\partial}{\partial r_{j}}U_{n}(x,y) = \frac{\zeta_{j}(t)}{m_{\rm{ion}}},
\end{equation}
where $U_{n}(x,y)$ represents the spatial dependence of the multipolar field, which can be written as
\begin{equation}
\label{eq:pot_even}
     U_{n}(x,y)=\sum_{k=0}^{m}\binom{2m}{2k}x^{2(m-k)}(-1)^{k}y^{2k},
\end{equation}
if $n$ is even ($n = 2m$ with $m$ $\epsilon \,\, \mathbb{N}^{+}$), or
\begin{equation}
     U_{n}(x,y)=\sum_{k=0}^{m}\binom{2m+1}{2k}x^{2(m-k) + 1}(-1)^{k}y^{2k},
\end{equation}
if $n$ is odd ($n = 2m + 1$ with $m$ $\epsilon \,\, \mathbb{N}^{+}$).

\begin{figure}[h]
\centering
\includegraphics[width=\columnwidth]{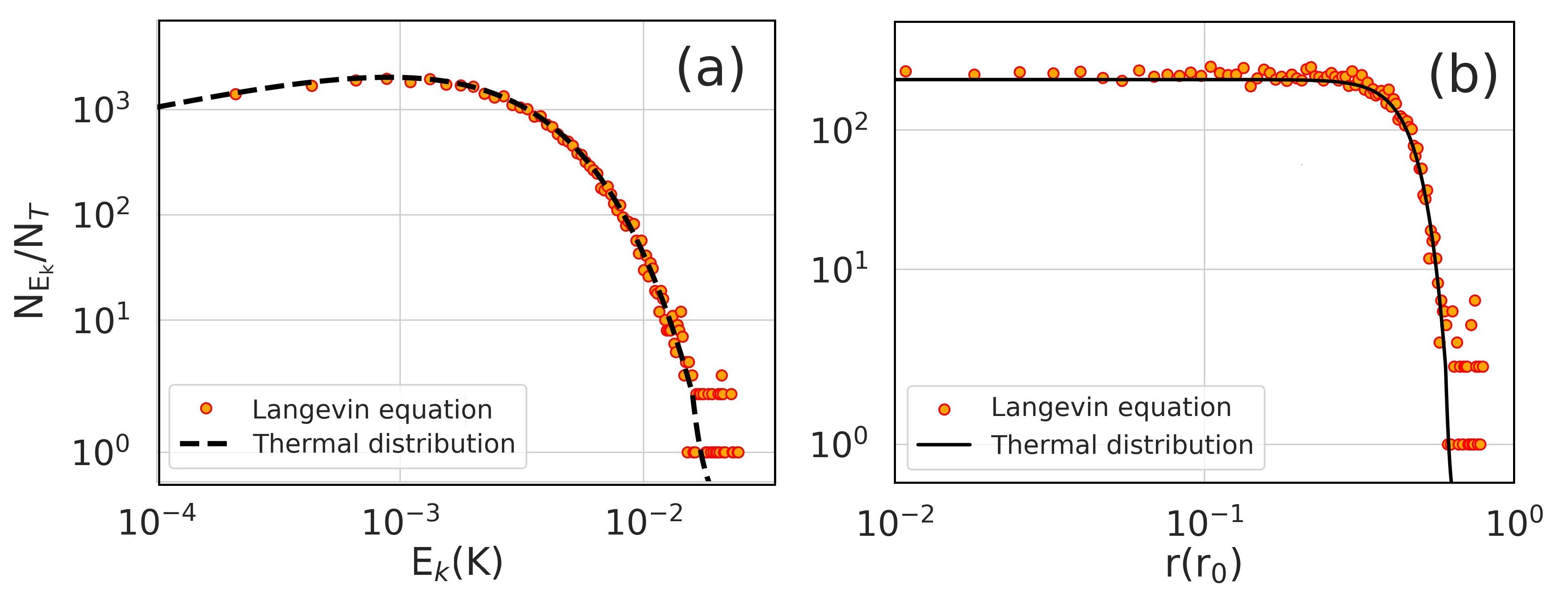}
\caption{Distributions obtained from the Langevin equation for (a) the ion's energy and (b) position, in a dodecapolar trap. Here the distribution does not depend on the atomic mass, but it does the relaxation time. The thermal distribution is displayed to verify the thermal behavior in this formulation.}
\label{fig:langevin_distributions}
\end{figure}

Eq.~(\ref{eq:langevin_eq}) represents a coupled stochastic differential equation for any n$>$2, in contrast to the case of a Paul trap~\cite{LONDONO2022}. Furthermore, Eq.~(\ref{eq:langevin_eq}) contains explicit time-dependent terms, and, as a result, there is no stationary solution for the associated Fokker-Planck equation. However, thanks to the Gaussian noise term (stochastic force) $\zeta_{j}$, the variables $r_{j}$ and $v_{i}$ will follow a time-averaged thermal distribution, as shown in Fig. \ref{fig:langevin_distributions}, where a comparison between the Langevin approach and the thermal distribution is displayed. The energy distribution of the ion is shown in panel (a) where it is noticed a wonderful agreement between our Langevin simulation and the thermal distribution. Similarly, in the case of the spacial distribution of the ion, panel (b), the Langevin formulation described the ion position extremely well in a thermal bath, given by Eq.~(\ref{eq:spatial_distribution}).

A stochastic formulation of the trapped ion dynamics in a neutral bath has allowed us to explore the continuous-time evolution of the physical quantities and, consequently, distributions, which is the primary advantage of a stochastic approach versus a molecular dynamics one. Here, we solve Eq.~(\ref{eq:langevin_eq}) using the leap-frog Verlet algorithm \cite{Risken1996} and take the average over 10$^{5}$ realizations of the ensemble to report on the mean time evolution of different quantities. We will use brackets to denote the ensemble average and the over bar for the time average.

\begin{figure}[h]
\centering
\includegraphics[width=\columnwidth]{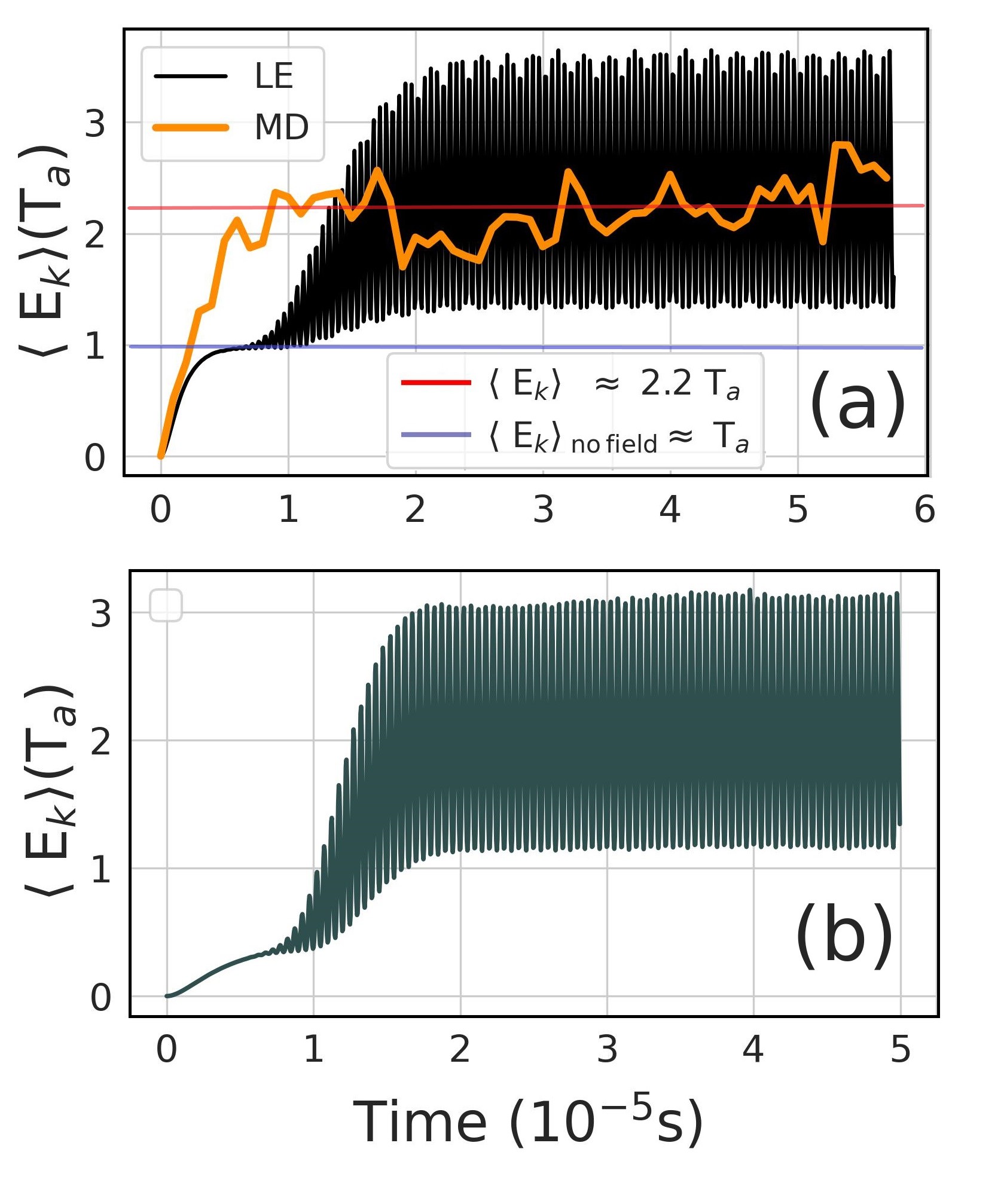}
\caption{Time evolution of the mean kinetic energy of Yb$^{+}$ in the presence of an Rb cloud at $T$ = 1$\times$10$^{-3}$K inside a dodecapolar ion trap (n = 6) using the Langevin equation model (LE). In panel (a) the sample density is $\rho$ = 1.0$\times$10$^{18}$m$^{-3}$ and the friction coefficient corresponds to $\gamma$ = 1.25$\times$10$^{-19}$ kg/s while for panel (b) we have $\rho$ = 7.2$\times$10$^{17}$m$^{-3}$ and $\gamma$ = 5.3$\times$10$^{-20}$ kg/s. For the shorter relaxation time (Panel (a)) we can see the double thermalization process resulting from the box-like potential nature of the n=6 trap, here we also show the evolution of the mean kinetic energy from the Molecular dynamics simulation (MD)}
\label{fig:avg_KE_evol}
\end{figure}

\begin{figure}[h]
\centering
\includegraphics[width=\columnwidth]{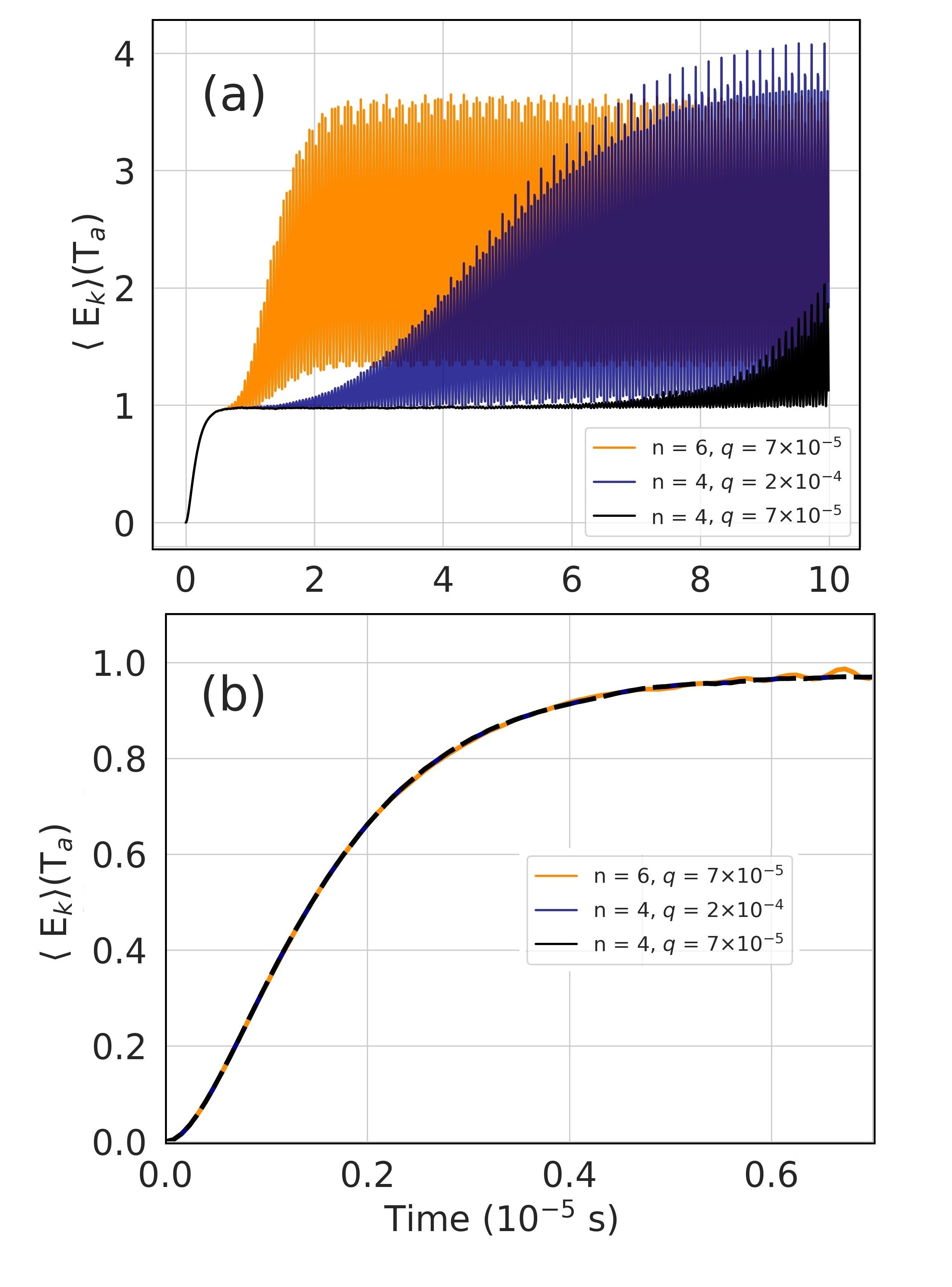}
\caption{Time evolution of the mean-kinetic energy  for different trap characteristics. Panel (a) shows the change in the second thermalization time for different trap properties, while panel (b) shows the thermal evolution, which is trap-independent. For all the simulations we use $T$=1$\times$10$^{-3}$K, $\rho$ = 1$\times$10$^{18}$m$^{-3}$ and $\gamma$ = 1.25$\times$10$^{-19}$kg/s  }
\label{fig:KE_for_q}
\end{figure}

Fig.~\ref{fig:avg_KE_evol} shows the evolution of the mean radial kinetic energy, $\langle E_{k} \rangle = \langle E_{k,x} \rangle + \langle E_{k,y}   \rangle $, for a single Yb$^{+}$ ion trapped in a dodecapolar trap ($n=6$) in the presence of a Rb cloud at $T = 1\times 10^{-3}$~K. In this Figure, it is noticed that the energy undergoes a double thermalization process. First, the ion energy thermalizes to a field-free value of k$_{\rm{B}}T$ according to the equipartition theorem. Then, once the potential starts to act significantly on the ion, energy is not a conserved quantity. The kinetic energy becomes a function of time and rapidly reaches a second time-ensemble average value, which is approximately equal to  2k$_{\rm{B}}T$, associated with the two micromotion degrees of freedom. The presence of a double thermalization process occurs if the relaxation time $\tau_{\rm{R}} = m/\gamma$ is lower than the time it takes for the ion to leave the field-free region of the trap. Therefore, the thermalization process is drastically affected by the density of the neutral bath, as one can see by comparing panels (a) and (b). Panel (a) shows a higher density baht than panel (b), and as a consequence, panel (a) shows abrupter thermalization dynamics in comparison with panel (b). Additionally, in panel (a), one also notices how the event-drive molecular dynamics simulation describes a similar tendency to the same mean energy value as in the Langevin equation model. However, the double thermalization is not observed due to the discrete steps in the simulation. Note that this observation validates the field-free approximation for the ion around the central region of the trap, absent in low order trap, in which subsequent thermalizations do not occur \cite{LONDONO2022}.

The time to reach the second thermalization value depends on the trap order and $q$, which is illustrated in Fig. \ref{fig:KE_for_q}. Increasing the $q$-parameter for the same trap order results in a lower critical radius, then, the effect of the field is felt by the ion at shorter distances, leading to shorter second-thermalization times, as shown in panel (a) of Fig.~\ref{fig:KE_for_q}. In the same panel, we notice that for the same $q$-parameter, a higher order trap leads to shorter thermalization time than in the case of low trap order, which is a consequence of the dependence of the effective potential amplitude with n$^{2}$. Finally, in Fig.\ref{fig:KE_for_q}(b) it is corroborated how the field-free dynamics is the same for each case as all of them have the same thermal and atomic properties of the bath.


\begin{figure}[h]
\centering
\includegraphics[scale=0.37
]{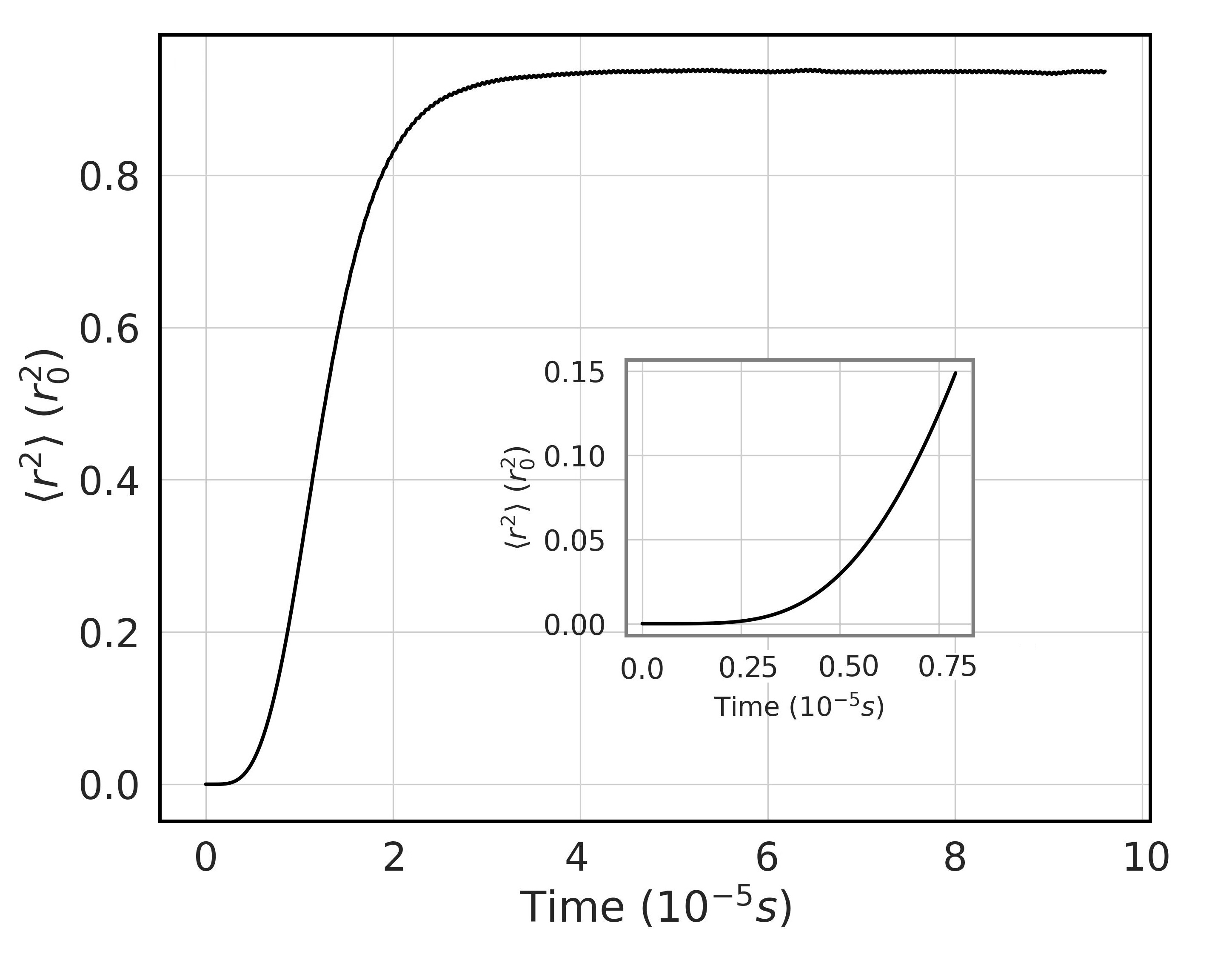}
\caption{Time evolution of the mean square radial displacement of Yb$^{+}$ in presence of a Rb cloud with density $\rho = 1\times 10^{18} m^{-3}$ and at $T$ = 1$\times$10$^{-4}$K inside a dodecapolar ion trap (n = 6) . The inside reveals the diffusive behavior at early times}
\label{fig:avg_rsmd}
\end{figure}

Another signature of the free-field evolution is a diffusive evolution at short times, characterized by a quadratic-to-linear evolution of the mean square radial displacement $\langle r^{2} \rangle$ = $\langle x^{2} \rangle$ + $\langle y^{2} \rangle$, as displayed in Fig.~\ref{fig:avg_rsmd}. On the contrary, at sufficiently long times, $\langle r^{2} \rangle$ saturates as characteristic of bounded stochastic motion. All these remarkable aspects of the time-continuous evolution of the ion's distributions result in a better understanding of the sympathetic cooling process and a helpful criterium for choosing traps to enhance the stability of the ion and its localization or control relaxation processes.



\section{Conclusions}
\label{sec:conclusions}
In this work, we have introduced two novel methodologies to simulate the dynamics of a single trapped ion in a multipolar trap immersed in a cold gas. First, we introduce the event-driven molecular dynamics simulation to explore the dynamic stability of the ion in this system. In addition, we develop a stochastic approach based on the Langevin equation. The first technique presents an analysis of the ion-bath dynamics in a multipole trap from a discrete-time perspective. In contrast, the stochastic approach leads to a continuous time description of the dynamics. 

The event-driven molecular dynamics is an ideal tool for trap stability studies. For example, the dodecapolar trap represents the optimal choice to reduce micromotion heating while the ion shows stable dynamics. Similarly, ion losses usually occur after thermalization for the range of considered temperatures, around T= 1$\times$10$^{-3} $K. In addition, thanks to the event-drive molecular dynamics simulations, we have derived an expression for the mean kinetic energy of the ion that generalizes previous attempts in the literature~\cite {BASTIAN2016}. On the other hand, the stochastic approach is a good fit for thermalization studies since it is a continuous-time approach and computationally cheap. Using this methodology, we predict a two-step thermalization mechanism of the ion. First, the ion thermalizes to the expected free-field case of $k_BT$, while in the second step, it reaches the expected $2k_BT$. 

Finally, these methods presented here are readily extensible to more involved experimental scenarios: including the excess of micromotion or the imperfections in electrodes. Therefore, these techniques could potentially impact the field of hybrid ion-atom system.

\section{Acknowledgments}
J.P.-R. thanks the Simons Foundation for the support. 

\newpage

\appendix

\section{Dynamic stability parameter}
\label{sec:appendixA}
In the case of multipole traps, for $\eta < \eta_{\rm{max}}$ the adiabatic approximation os valid~\cite{DIETER1992}, and in the absence of a DC voltage, the multipole trap potential is given by 

\begin{equation*}
   V_{\rm{RF,n}} = \frac{U_{\rm{AC}}}{r_{0}^{n}}\cos(\Omega_{\rm{RF}}t)r^{n}.
\end{equation*}
Therefore, the ion's equation of motion reads as
\begin{equation}
\label{eq:ion_EOM}
    m_{i}\frac{d^{2}\textbf{r}(t)}{dt^{2}} = \textbf{F}(\textbf{r}) = \frac{eU_{\rm{AC}}n}{r_{0}^{n}}\cos(\Omega_{\rm{RF}}t)r^{n-1}\textbf{e}_{\rm{RF}},
\end{equation} 
where $\textbf{e}_{\rm{RF}}$ represents the unitary vector of the trapping force, which depends on the azimuthal angle $\phi$. 

Let us assume that within the adiabatic approximation the position of the ion can be written as 
 $\textbf{r}(t) = \textbf{r}_{\text{sec}}(t) + \textbf{r}_{\text{mm}}(t)$, where $\textbf{r}_{\text{sec}}$ and $\textbf{r}_{\text{mm}}$ correspond to the secular and micromotion components of the ion's motion, respectively, satisfying $|\textbf{r}_{\text{sec}}| \gg |\textbf{r}_{\text{mm}}|$
and $|\ddot{\textbf{r}}_{\text{sec}}| \ll |\ddot{\textbf{r}}_{\text{mm}}|$. Then, Eq.(\ref{eq:ion_EOM}) reads as

\begin{equation}
\label{eq:eom_adiabatic}
    m_{i}\frac{d^{2}}{dt^{2}}(\textbf{r}_{\text{sec}}(t) + \textbf{r}_{\text{mm}}(t)  ) \approx \textbf{F}(\textbf{r}_{\text{sec}}) + (\textbf{r}_{\text{mm}}\nabla)\textbf{F}(\textbf{r}_{\text{sec}}),
\end{equation}
where a Taylor expansion has been performed on the force term up to first order in $r_{mm}$. 
Equating the most relevant terms in both sides of the equation, we find the differential equation for the micromotion as

\begin{equation}
\begin{split}
        \frac{d^{2}}{dt^{2}} \textbf{r}_{\text{mm}} = \textbf{F}(\textbf{r}_{\text{sec}}) &= -\frac{eU_{\text{AC}}n}{{m_{i}}r_{0}^{n}}\cos(\Omega_{\rm{RF}} t)r_{\rm{sec}}^{n-1}\textbf{e}_{\rm{RF}},
\end{split}
\end{equation}
that is easily solvable taking into account that the micromotion and secular are decoupled leading to 
\begin{equation}
    \textbf{r}_{\text{mm}} \approx \frac{eU_{\text{AC}}n}{{m_{i}r_{0}^{n}}\Omega_{\rm{RF}}^{2}}\cos(\Omega_{\rm{RF}} t)r^{n-1}_{\rm{sec}}\textbf{e}_{\rm{RF}}.
\end{equation}
Using this approximate solution we find the equation for the secular motion as
\begin{equation}
\begin{split}
    \frac{d^{2}}{dt^{2}}\textbf{r}_{\text{sec}} &= \frac{1}{m_{i}}(\textbf{r}_{\text{mm}}\nabla)\textbf{F}(\textbf{r}_{\text{sec}}) \\& \approx \frac{e^{2}U_{\text{AC}}^{2}n^{2}}{m_{i}^{2}r_{0}^{2n}\Omega^{2}}(n-1)\cos^{2}(\Omega_{\rm{RF}} t)r_{\rm{sec}}^{2n-3}\textbf{e}_{\rm{r}},
    \end{split}
\end{equation}
where $\textbf{e}_{\rm{r}}$ is the radial unitary vector. Averaging over the fast RF oscillations we see that the secular component follows the equation
\begin{equation}
\label{eq:secular_motion}
    \frac{d^{2}}{dt^{2}}\textbf{r}_{\text{sec}} = \frac{q^{2}n^{2}\Omega^{2}}{4}(n-1)\frac{r_{\rm{sec}}^{2n-3}}{r_{0}^{2n-4}}
\end{equation}
with $q$ = $\frac{2eU_{\text{AC}}}{r_{0}^{2}m_{i}\Omega_{\rm{RF}}^{2}}$. 

Equation \ref{eq:secular_motion} represents the periodic motion of the ion in the pseudopotential
\begin{equation}
\label{eq:pseudopotential}
    V_{\rm{eff}}(r)= m_{\rm{ion}}\frac{q^{2}n^{2}\Omega^{2}}{16}\frac{r_{\rm{sec}}^{2n-2}}{r_{0}^{2n-4}},
\end{equation}
generated from averaging the energy associated with the micromotion.

This adiabatic decomposition of the ion's motion is valid always we keep the conditions $|\textbf{r}_{\text{sec}}| \gg |\textbf{r}_{\text{mm}}|$
and $|\ddot{\textbf{r}}_{\text{sec}}| \ll |\ddot{\textbf{r}}_{\text{mm}}|$ as mentioned previously. The validity of this condition is evaluated through the quotient between the two terms on the right-hand side of equation \ref{eq:eom_adiabatic}. This leads us to define the stability parameter 
\cite{BASTIAN2016}
    \begin{equation}
    \eta  = qn(n-1)\bigg( \frac{r}{r_{0}} \bigg)^{n-2}.
\end{equation}

\section{ Multipolar potential in Cartesian coordinates}
\label{sec:appendixB}
The previous formulations can be also described in terms of cartesian coordinates. This version could result more convenient for some numerical aplications as in the case of the Langevin equation described in Sec.(\ref{sec:langevin_eq}). Then, it is imperative to writte the spatial part of the multipolar potential in cartesian coordinates as shown here.

The radial part of this potential can be written as
\begin{equation}
\label{eq:multipolar_potential}
    V_{\text{RF},n}(x,y,t) = (U_{\text{DC}}+ U_{\text{AC}}\cos(\Omega_{\rm{RF}} t))U_{n}(x,y),
\end{equation}
where $U_{\text{DC}}$ and $U_{\text{AC}}$ refers to the continuum and alternating potential amplitude in the electrodes
and $U_{n}(x,y)$ is the spatial dependence of the potential.  

$U_{n}(x,y)$ is a harmonic function and it can be built up, in the ideal electrode case, from the real part of the analytical complex function \cite{Friedman1982}
\begin{equation*}
    z^{n} = (x+iy)^{n} = U_{n}(x,y) + iV_{n}(x,y).
\end{equation*}

If $n$ is even ($n = 2m$ with $m$ $\epsilon \,\, \mathbb{N}^{+}$), we can write the potential as
\begin{equation*}
    \begin{split}
        U_{n}(x,y)=\sum_{k=0}^{m}\binom{2m}{2k}x^{2(m-k)}(-1)^{k}y^{2k},
    \end{split}
\end{equation*}
$x$ and $y$ are going to have the same even exponents between 0 and $n$. From this potential, we derived the spatial part of the force components
\begin{equation}
    \begin{split}
        \frac{\partial U_{n}(x,y)}{\partial x} &=\sum_{k=0}^{m-1}\binom{2m}{2k}2(m-k)x^{2(m-k)-1}(-1)^{k}y^{2k}\\
        & = 2m\sum_{k=0}^{m-1} \binom{2m-1}{2k}x^{2(m-k)-1}(-1)^{k}y^{2k},
    \end{split}
\end{equation}
and 
\begin{equation}
    \begin{split}
        \frac{\partial U_{n}(x,y)}{\partial y} &=\sum_{k=1}^{m}\binom{2m}{2k}2(k)x^{2(m-k)-1}(-1)^{k}y^{2k-1}\\
        & = 2m\sum_{k=1}^{m} \binom{2m-1}{2k-1}x^{2(m-k)}(-1)^{k}y^{2k-1}.
    \end{split}
\end{equation}
Then both of the component of the force have the same number of terms with the same binomial coefficients. 

For an odd-$n$ trap, ($n = 2m+1$ with $m$ $\epsilon \,\, \mathbb{N}^{+}$), the spatial dependence of the forces take the form
\begin{equation}
    \begin{split}
        \frac{\partial U_{n}(x,y)}{\partial x} 
        & = (2m+1)\sum_{k=0}^{m} \binom{2m}{2k}x^{2(m-k)}(-1)^{k}y^{2k},
    \end{split}
\end{equation}
and 
\begin{equation}
    \begin{split}
        \frac{\partial U_{n}(x,y)}{\partial y} 
        & = (2m+1)\sum_{k=1}^{m} \binom{2m}{2k-1}x^{2(m-k)+1}(-1)^{k}y^{2k-1}.
    \end{split}
\end{equation}

and the components do not have either, the same number of terms nor the same binomial coefficients. This results in a remarkable difference between the $x$ and $y$ dynamics.

\section{Hamonic contribution to $\langle r^{2} \rangle$ and $\langle v^{2} \rangle$ }

\begin{figure}[htbp]
\centering
\hspace{-0.1cm}
\includegraphics[scale=0.52
]{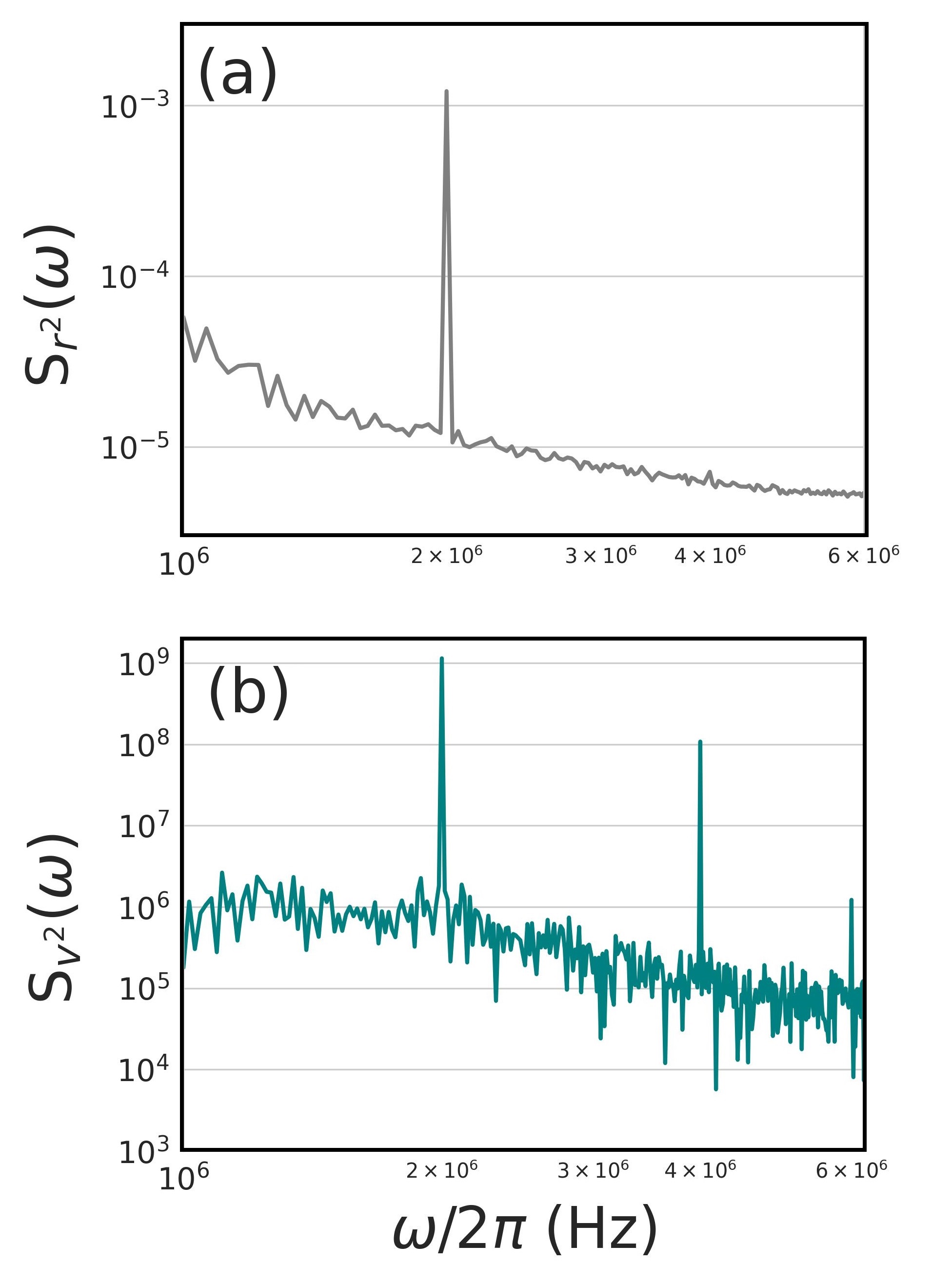}
\caption{Power spectrum of the $\langle r^{2} \rangle$ and $\langle v^{2} \rangle$ evolution for a Yb$^{+}$ ion in the presence of an atomic cloud of Rb with $T$=1$\times$10$^{-3}$K, $\rho$ = 1$\times$10$^{18}$m$^{-3}$ and $\gamma$ = 1.25$\times$10$^{-19}$kg/s . The peaks are located at the even harmonics of the trap frequency and the noise comes from the no averaged thermal fluctuations}
\label{fig:spectral_density}
\end{figure}

Additional dynamical aspects can be studied from the Langevin dynamics methodology, which brings light to some questions for instance the ion's localization. Here we address the harmonic behavior of the mean square velocity and position of the ion. Fig.~\ref{fig:spectral_density} shows the power spectrum for the evolution of $\langle r^{2} \rangle$ and $\langle v^{2} \rangle$. From this plot, two main aspects can be highlighted: First, the remarkable difference between the amplitude of the radio-frequency oscillations in the evolution of $\langle v^{2} \rangle$ and $\langle r^{2} \rangle$. The amplitude of the oscillations in $\langle r^{2} \rangle$ is so small compared to its mean value that it can be approximated to a time-independent variable. Second, we notice that the evolution of $\langle v^{2} \rangle$ contains more harmonic contributions from the fundamental trap frequency $\Omega_{\rm{RF}}$ than  $\langle r^{2} \rangle$. Furthermore, only even harmonic contributions appear. These two properties are only characteristic of traps where $n = 2m$, and $m$ is an odd number, as is the case of the dodecapolar trap (n = 6, m = 3).

To understand why this happened we can start noticing that the spatial part of the trapping force for $x$ and $y$ components are, following eq. \ref{eq:pot_even}, 
\begin{equation}
\begin{split}
    \label{eq:forces_x_and_y}
    &\frac{\partial }{\partial x}U_{n}(x,y) = \sum_{k=0}^{m-1} 2(m-k) \binom{2m}{2k}x^{2(m-k)-1}(-1)^{k}y^{2k}\\
    &\frac{\partial }{\partial y}U_{n}(x,y) = \sum_{k=0}^{m-1} 2k \binom{2m}{2k}x^{2(m-k)}(-1)^{k}y^{2k-1},
\end{split}    
\end{equation}
respectively. Further manipulation of the $y$-component led us to the expression
\begin{equation}
\begin{split}
    \label{eq:forces_y}
    \frac{\partial }{\partial y}U_{n}(x,y) = (-1)^{m}\sum_{k=0}^{m-1} 2(m-k) \binom{2m}{2k}y^{2(m-k)-1}(-1)^{k}x^{2k},
\end{split}    
\end{equation}
which is exactly the $x$-component but with the change $x \rightarrow y$ and the front sign $(-1)^{m}$.

\begin{figure}[htbp]
\centering
\hspace{-0.1cm}
\includegraphics[scale=0.34
]{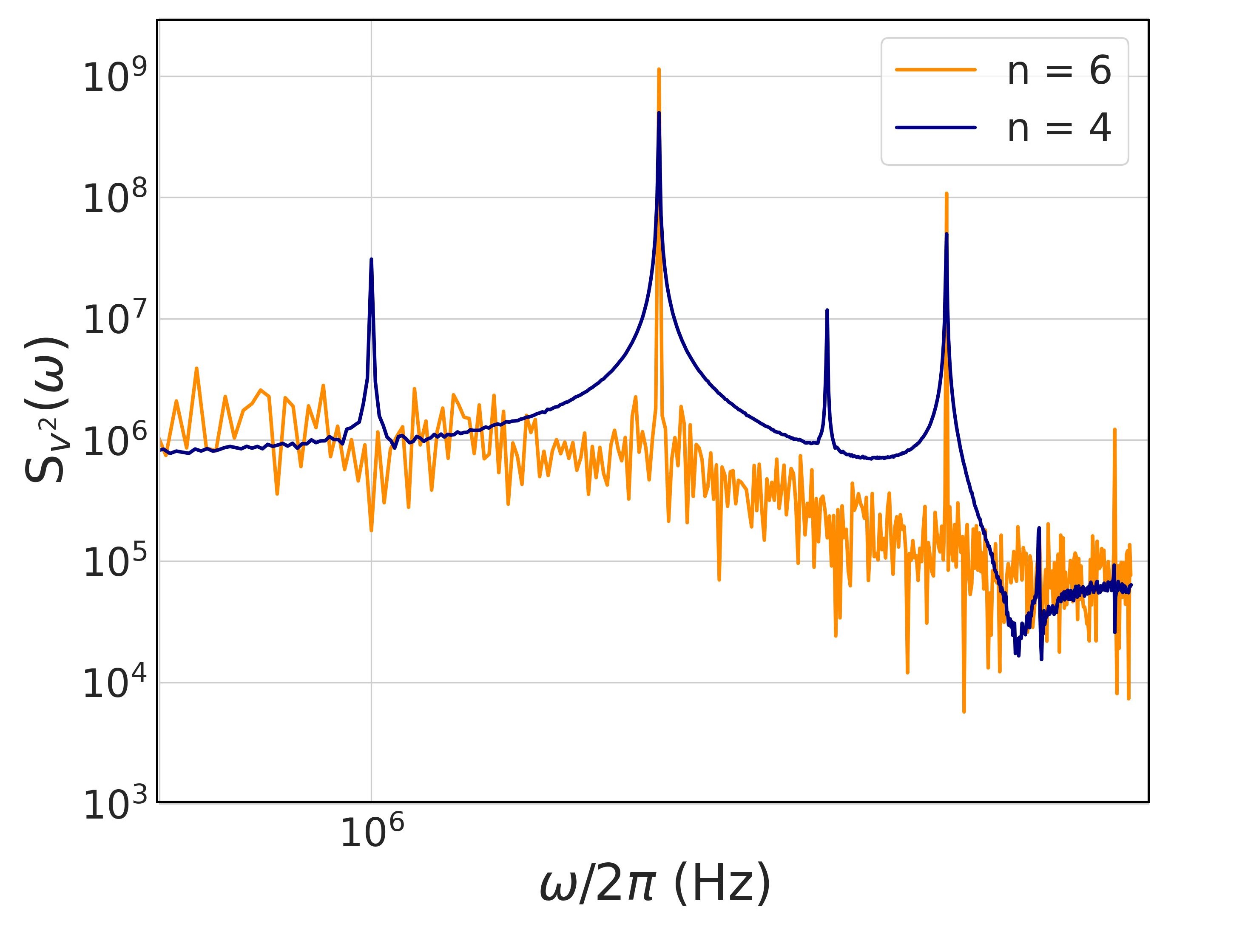}
\caption{Power spectrum of the evolution of $\langle v^{2} \rangle$ for two different multipole traps with the same parameters as in Fig. \ref{fig:spectral_density}. It can be seen that for n=4 (m=2) even and odd harmonic contributions of the trap frequency appear, while for n=6 (m=3) only odd contributions appear. The n=4 case has been averaged over 1E6 realizations to further avoid thermal noise.}
\label{fig:spectral_density_n6_vs_n4}
\end{figure}

Now, for long times ($t \gg  \tau_{\rm{c}}$ ) we can express the solution for the mean square value of each position component as a Fourier series \cite{LONDONO2022, BLATT1986}
\begin{equation}
    \langle r_{j}^{2} \rangle = \sum_{n}r_{j,n}e^{-i n\Omega_{\rm{RF}}t},
\end{equation}
where the Fourier coefficients $r_{j,n}$ are going to depend, among other things, on the n-th power of the $q$-parameter \cite{LONDONO2022, Drewsen2000}. Then, if m is odd, the spatial part of the trapping force is going to be the same for $x$ and $y$ components but with the opposite sign because of the (-1)$^{m}$ term (see eqs. \ref{eq:forces_x_and_y} and \ref{eq:forces_y}). We can assign this different sign to the $q$  factor as usual in the linear Paul trap such that $q_{x} = -q_{y}$, doing this, all the Fourier coefficients become identical for $\langle x^{2} \rangle$ and $\langle y^{2} \rangle$ but with a different sign which is only manifest in the odd powers of the $q$-factor, it means, a negative sign is going to accomplish the odd harmonic contribution for the $y$ component. As a consequence, when defining $\langle r^{2} \rangle $ for long times we have 
\begin{equation}
    \begin{split}
        \langle r^{2} \rangle &=  \langle x^{2} \rangle +  \langle y^{2} \rangle \\
        &= (x_{0} + x_{1}e^{-i \Omega_{\rm{RF}}t} +...) + (x_{0} - x_{1}e^{-i \Omega_{\rm{RF}}t} +...)\\
        & = 2x_{0} + 2x_{2}e^{-i \Omega_{2\rm{RF}}t}+... = 2\sum_{n} x_{2n}e^{-2in\Omega_{\rm{RF}}t}.
    \end{split}
    \label{eq:mean_sq_position}
\end{equation}
So, the first time-dependent contribution is second order in $q$, which is small for most of the stable configurations found in Sec. \ref{sec:stability}.
In general, this time independence of the mean square radial displacement results in better localization properties of the ion inside the trap. We should also notice that high-order contributions of the harmonics for the mean square velocity are larger than the one for the mean square position, which results in the strong time-dependence of $\langle v^{2} \rangle$. The same arguments result in eq. \ref{eq:mean_sq_position} keeps for $\langle v^{2} \rangle$. In fig. \ref{fig:spectral_density_n6_vs_n4} we show the power spectrum of $\langle v^{2} \rangle$ for the n=6 and n=4 traps. The octupolar trap shows three additional peaks in the spectrum, one at the fundamental RF-frequency and the other two at the third and fifth harmonic, verifying the previous analysis. 

\newpage

\bibliography{apssamp}


\end{document}